\documentstyle{article}

\def\comm#1#2{\left[#1 , #2 \right]}
\def\pss#1#2{ \left\{ #1, #2 \right\}_{PB} } 
\def\cmm#1#2{ \left[ #1, #2 \right] }   
\def\acm#1#2{ \left\{ #1, #2 \right\} } 

\begin{document}

\begin{center}

{\Large \bf
 Deformed oscillator algebras
 for
 two--dimensional
 quantum superintegrable systems}\\

\bigskip
{Dennis Bonatsos $^1$, C. Daskaloyannis $^2$ and K. Kokkotas $^2$ }

\medskip
{$^1$ Institute of Nuclear Physics, N.C.S.R. ``Demokritos'',
GR--15310 Aghia Paraskevi, Attiki, Greece}

{$^2$ Department of Physics, Aristotle University of
Thessaloniki, GR--54006 Thessaloniki, Greece }
\bigskip

 \end{center}

\vskip 0.5in
\begin{center}
{\bf Abstract}
\end{center}

Quantum superintegrable systems in two dimensions are obtained from
their classical counterparts, the quantum integrals of motion being
obtained from the corresponding classical integrals by a symmetrization
 procedure.  For each quantum superintegrable system
a deformed oscillator algebra, characterized by a structure function
specific for each system, is constructed, the generators of the algebra
being functions of the quantum integrals of motion. The energy
eigenvalues  corresponding to a  state with finite
dimensional degeneracy
can then be obtained in an economical way from solving a system
of two equations satisfied by the structure function, the results being
in agreement to the ones obtained from the solution of the relevant
Schr\"{o}dinger equation. Applications to the harmonic oscillator in
a flat space and in a curved space with constant curvature,
 the Kepler problem  in a flat or curved space, the
Fokas--Lagerstrom potential, the Smorodinsky--Winternitz potential,
and the Holt potential are given. The method shows how quantum
algebraic techniques can simplify the study of quantum superintegrable
systems, especially in higher dimensions.

\vskip 0.3in

\section{Introduction}

The idea of studying the properties of physical systems exhibiting
degenerate energy levels through use of
their symmetries has been exploited since the early days of quantum mechanics.
In these cases the symmetry algebra \cite{MQL}
of the system has to be determined, which is a
finite-dimensional Lie algebra containing ladder operators which connect
all the eigenstates with a given energy, while the Hamiltonian of the
system is related to the Casimir operators of the algebra. The set of
eigenstates with given energy provides a basis for an irreducible
representation (irrep) of the algebra. The energy eigenvalues are then
determined by the eigenvalues of the Casimir operators of the algebra
in the corresponding irreps.
The determination of the symmetry algebra of a given hamiltonian is a quite
difficult task, which is not always dictated by an a priori obvious
procedure of searching for it. Examples of physical systems
having known symmetry algebras  are the N-dimensional isotropic
harmonic oscillator and the Kepler (Coulomb) system, bearing the symmetries
SU(N) and SO(N+1) respectively.

The classical counterparts of the quantum isotropic harmonic oscillator
and the quantum   Kepler problem have
another interesting property. They are maximally superintegrable
systems in N dimensions. Superintegrable systems in N dimensions have
more than N independent classical constants (also called integrals or
invariants) of motion, while maximally superintegrable systems in
N dimensions   have 2N$-$1 independent
classical constants of motion, N of which
 are integrals in involution (the Poisson
bracket of each pair of them is zero).
This property of the maximally superintegrable systems
implies, in classical mechanics, that every closed
and bounded trajectory is a periodic trajectory.  A detailed review of
superintegrable systems in 2 dimensions is given in \cite{Hiet1}, while
examples of superintegrable systems in 3 dimensions can be found in
\cite{Ev,KW1}.

Higgs \cite{Higgs} and Leemon \cite{Leemon} have shown that in
the case of a N dimensional system moving in a space with
constant curvature the isotropic harmonic oscillator and the Kepler
problem are still maximally superintegrable systems, both in classical
and quantum mechanics. Furthermore, they have shown that the
quantum counterparts of these systems
can be described by symmetry algebras
isomorphic to  SU(N) and SO(N+1) respectively.
Additional examples of superintegrable classical systems are
the Fokas--Lagerstrom potential \cite{FL},
the Smorodinsky--Winternitz potential \cite{SW1,SW2,SW3,Ev2},
the Holt potential \cite{Holt},
the Hartmann potentials \cite{H1,H2,H3,KW2}. The problem of quantum
integrability and its connection to classical integrability
is currently under active investigation \cite{Weigert,Robnik,Korsch,Hiet2}.

In several of the above mentioned cases (see \cite{Higgs,Leemon}, for example)
the property of the classical and quantum
superintegrability of a physical system coincides   with the existence of
a symmetry algebra of the system. The energy levels of the system can
then be determined by purely algebraic means. However, the identification of a
symmetry algebra is not always easy. Furthermore, in some cases (see
\cite{KW2}, for example) it seems that the usual Lie algebras do not
suffice for this purpose. The recent introduction of quantum algebras
\cite{KR,Sklyanin,Drinfeld,Jimbo} (also called quantum groups) opens
new possibilities in this direction. Quantum algebras are nonlinear
deformations of the usual Lie algebras, to which they reduce when the
deformation parameter $q$ goes to 1. They were  initiated as a
mathematical tool issued from the study of the quantum
inverse problem, the Yang Baxter equation and conformal field theories
(see \cite{Jimbo2} for a collection of original papers). There are
already some indications that quantum algebras might be useful as
symmetry algebras of certain superintegrable systems.
The Higgs algebra (i.e. the symmetry algebra of the Kepler problem in a
2-dimensional space with constant curvature studied in \cite{Higgs})
can be approximated to second order by the quantum algebra SU$_q$(2)
\cite{Z1}. The symmetry algebra of the Hartmann potential, for which
usual Lie algebras seemed insufficient \cite{KW2}, has been identified
as the quadratic Hahn algebra QH(3) \cite{Z2}. The quadratic Hahn algebra
QH(3) has also been found to describe the symmetry of the anisotropic
singular oscillator (a 3-dimensional
harmonic oscillator with an additional term
$\sim 1/(r^2 \sin^2\theta)$ \cite{Z3}.
The quadratic Hahn algebra QH(3) is a special case of the
general quadratic Askey--Wilson algebra QAW(3), which is the dynamical
symmetry of the potentials having eigenfunctions described by
classical polynomials \cite{Z4}.
L\'{e}tourneau and Vinet \cite{LV}  have constructed
the quadratic algebra describing the case of a harmonic oscillator
potential with a 2 to 1 frequency ratio.
 Another example of a system
described by a nonlinear algebra is the generalized Kepler (Coulomb)
system \cite{KC}. From these examples it becomes clear that nonlinear
algebras can be useful in the description of integrable and superintegrable
systems.

In this paper we focus attention on the simplest superintegrable systems,
 the two dimensional ones, and we propose a method of determining
their  dynamical symmetries and calculating their spectra by
purely algebraic means. It turns out that several quantum superintegrable
systems in 2 dimensions can be described in terms of appropriate
generalized deformed oscillators, which allow for the direct determination
of the energy levels and their degeneracies
 without any need of solving the Schr\"odinger
equation.
 A preliminary study of the proposed method can be found in
ref. \cite{PRAnote}, where the method has been used in two cases of potentials,
the symmetric harmonic oscillator in a curved space and
the asymmetric oscillator with a 2 to 1 frequency ratio. In this
paper we study most of the known
2 dimensional superintegrable cases.

It is known that
the classical (non-deformed) algebras can be constructed using
the harmonic oscillator algebra $\{ a, a^+, N\}$ as the
basic underlying structure. For quantum algebras, and nonlinear
algebras in general, deformed oscillators have to be used.
Biedenharn \cite{Biedenharn} and
Macfarlane \cite{Macfarlane} constructed the q-deformed oscillator
appropriate for the Schwinger realization of the quantum algebra
 SU$_q$(2). Many other deformed oscillators can be found in the
literature. We mention the Q-oscillator introduced by Arik and
Coon \cite{AC} and Kuryshkin \cite{Ku}, the two-parameter deformed
oscillator \cite{JBM,CJ}, the parafermionic oscillator \cite{OK}
and its q-deformation \cite{FV,OKK}, the parabosonic oscillator
\cite{OK} and its q-deformation \cite{FV,OKK}, the generalized
q-deformed fermionic algebra \cite{VPJ}. (The q-deformed fermionic
algebra \cite{Hayashi} has been proved to be equivalent to the usual
fermionic algebra \cite{BD1}.)
The common feature of all these deformations is their structural  similarity.
In all cases  an appropriate Fock
basis can be constructed, leading to a matrix representation of
the algebra.

The structural similarity of the various deformed oscillators implies
that all of them can be described in a unified framework. Among the
various alternatives we mention here
the generalized deformed oscillator \cite{D1},
the Odaka--Kishi--Kamefuchi unification scheme \cite{OKK},
the pioneering work of Jannussis {\it et al.} on the bozonization
method  \cite{JBSZ} and the generalized Q-deformed oscillator
\cite{BJ},
the Beckers--Debergh unification scheme \cite{BD}, and
the Fibonacci oscillator scheme \cite{AD}, while  a general
treatment of qualgebras is given recently by Fairlie and Nuyts
\cite{FN}.

 Among the various equivalent descriptions of the
deformed oscillators, in this paper we use the deformed
oscillator algebra \cite{D1}, already used for the study of
the energy spectra of one dimensional systems \cite{D2,D3,BD2}.

In section 2 of the present paper we consider the classical superintegrable
systems in 2 dimensions, for which we show that
 the relevant Poisson bracket induced algebra has a
structure similar to the deformed oscillator algebra.
In section 3 a working hypothesis for the quantum superintegrable systems
in 2 dimensions is proposed,  which leads  to the
calculation of the energy eigenvalues and their degeneracies
 by purely algebraic means.
This hypothesis is applied to several quantum superintegrable
examples in section 4, while section 5 contains discussion of the
present results and plans for further work.

\section{Classical superintegrable systems in two dimensions}

Consider a classical system with two degrees of freedom, described by
the Hamiltonian:
\begin{equation}
H=H(x,y,p_x,p_y).
\label{eq:hamiltonian}
\end{equation}
If the system is superintegrable there are two independent additional
integrals of motion $I$ and $C$, such that:
\begin{equation}
\pss{H}{I}=\pss{H}{C}=0, \quad\mbox{and}\quad \pss{I}{C}=
F(H,I,C),
\label{eq:PoissonRelations}
\end{equation}
where $ \pss{\ }{\ }$ denotes the Poisson bracket and
 $F=F(H,I,C)$ is a constant of motion which depends on the
three independent constants of motion $H,I,C$.
A superintegrable system in two dimensions is necessarily a maximally
superintegrable system, which means that  all finite classical trajectories
are closed and  periodic. Integrable and superintegrable systems in 2
dimensions have been reviewed in \cite{Hiet1}, while in \cite{Ev}
a systematic study is given of superintegrable systems in 3 dimensions
which possess invariants that are quadratic polynomials of the canonical
momenta.

Maximally superintegrable systems possess, by definition,  the maximum
number of independent classical invariants. Therefore
any other integral can be expressed as a function of the basic
integrals $H$, $I$, $C$. As a result we can in general choose two new
integrals of motion:
$$ L=L(H,I,C), \quad \mbox{and} \quad A=A(H,I,C), $$
such that:
\begin{equation}
\pss{L}{A}= B, \quad \pss{L}{B}=-A.
\label{eq:Cl1}
\end{equation}
After a calculation we can prove that:
$$
B^2+A^2=G(H,L),
$$
where $G(H,L)$ is some function depending only on the integrals
of motion $H$, $L$, and
\begin{equation}
\pss{A}{B}=\Phi (H,L)=- \frac{1}{2} \frac{\partial G}{\partial L}.
\label{eq:Cl2}
\end{equation}

The structure of the algebra defined by eqs
(\ref{eq:Cl1}-\ref{eq:Cl2})  has many similarities  to the algebraic
structure of the deformed oscillator given in references
\cite{D1,D2,D3}, where $L$ is some kind of number operator, while $A$, $B$ are
like the creation and annihilation operators. The deformed oscillator
algebra is a non abelian algebra.
Therefore it is quite natural to attempt
studying the quantum superintegrable systems by applying some
similar procedure in order to calculate their
 quantum properties (eigenvalues and eigenvectors), since the
corresponding properties of the deformed oscillators have been
 already studied.

\section{ Quantum superintegrable systems in two dimensions}

The question of integrability in quantum mechanics is under
 investigation. Many authors
\cite{Weigert,Robnik,Korsch,Hiet2}
 have investigated several  aspects
of the extension of integrability from classical to
quantum mechanics. We should recall at this point
 that each quantum system with
discrete energy spectrum can be considered as a quantum
integrable system \cite{Weigert}.

In this paper we consider a two dimensional quantum system
described by a hamiltonian $H$ acting on a Hilbert space $\cal H$.
The hamiltonian is an autoadjoint operator with range dense in the
space $\cal H$.
All the operators defined in this section are supposed to be
generated by nonlinear combinations of the basic algebra of
generators $x,p_x,y,p_y$ satisfying the usual commutation
relations:
$$
\comm{x}{p_x}=\comm{y}{p_y}=i, \quad \mbox{other commutators =
0} .
$$

 If the system is integrable, then there is a
second autoadjoint operator $I$ commuting with the hamiltonian
and having a range dense in $\cal H$
\begin{equation}
\comm{H}{I}=0,
\label{eq:comm}
\end{equation}
the operators $H$ and $I$ being linearly independent.
The commutativity of these operators implies that
there is a family of common eigenvectors for both operators.
Let us label the common eigenvectors of these operators  using
their corresponding eigenvalues, thus obtaining a family
of vectors in the Hilbert space ${\cal H}$.
Let us further suppose that this family spans the whole Hilbert space
 and that there is no simultaneous degeneracy for
both labels of the common eigenfunctions
(i.e. there could be degeneracy in each label separately, but not in both
labels simultaneously). This assumption defines  some
kind of {\em completeness} of the system of the two quantum integrals
$H,I$ in involution. The present assumption  is also consistent
with the fact that the ranges of the
operators $H$ and $I$ are supposed to be dense in the ambient
Hilbert space $\cal H$.

Weigert \cite{Weigert} has proved that there is another quantum
integral $I^\prime$ possessing the same properties as $I$. This is
permissible in quantum mechanics but not in classical mechanics.
 In this sense the choice
of complete operators $H$, $I^\prime$ instead of $H$, $I$
 means a different way of labelling of
the base spanning the Hilbert space.
Here we shall treat the part of the Hilbert space corresponding to
 a discrete spectrum. In some cases the whole of the
Hilbert space can be described by a discrete basis (as in the
case of the harmonic oscillator), while in other cases (as in the
case of the Coulomb (Kepler) potential) a part of the Hilbert
space is described by a discrete basis (the space of energy
eigenvectors with negative energy eigenvalues).

  The
system is called {\em superintegrable}, by analogy to the classical
definitions, if there is a third operator $C$, linearly
independent from $H$ and $I$,  with range dense in the
Hilbert space and commuting with $H$ but {\bf not} commuting
with $I$
$$
\comm{H}{C}=0,\quad \comm{I}{C}\ne 0.
$$

In this paper we propose the following working hypothesis:

{\bf Hypothesis}:
 {\em Let us consider the superintegrable systems for which we
can construct an associative  algebra:
\begin{equation}
\begin{array}{rl}
{\cal N} =& {\cal N} \left( H, I ,C \right), \\
{\cal N}^+ =& {\cal N},\\
{\cal A} =& {\cal A} \left( H, I ,C \right), \\
\comm{\cal N}{\cal A}=& - {\cal A},\\
{\cal A}^+ {\cal A} =& \Phi \left( H,{\cal N} \right),\\
\comm{ {\cal A}^+ {\cal A}}{ {\cal A} {\cal A}^+}=&0,
\end{array}
\label{eq:hypothesis}
\end{equation}
where $\Phi (E,x)$ is a real positive function definite for $x \ge
0$ and
 \begin{equation}
\Phi (E,0)=0.
\label{eq:condition1}
\end{equation}
}

{}From the above equations we can prove that:
$$
\begin{array}{rl}
\comm{\cal N}{{\cal A}^+}=&  {\cal A}^+, \\
{\cal A} {\cal A}^+ =& \Phi \left(H, {\cal N} + 1 \right).
\end{array}
$$
If this construction is possible we can then define the Fock space
for each energy eigenvalue:
$$
\begin{array}{rl}
H\vert E,n>=&E\vert E,n>, \\
{\cal N} \vert E, n> = &n \vert E,n>, \; n=0,1\ldots, \\
{\cal A} \vert E, 0>=&0, \\
\vert E, n>=& \left( \frac{1}{\sqrt{[n]!}}\right)
 \left( {\cal A}^+ \right)^n  \vert E, 0>,
\end{array}
$$
where
$$
[0]!=1, \quad [n]!=\Phi(E,n)[n-1]! .
$$
In the case of the discrete energy eigenvalues, for every energy
eigenvalue $E$ there is some degeneracy of dimension $N_d+1$.
Therefore the dimensionality of the Fock space corresponding to
that energy eigenfunction should be equal to $N_d+1$. This is
equivalent to the condition:
\begin{equation}
\Phi (E,N_d+1)=0.
\label{eq:condition2}
\end{equation}
As we shall see in the examples given in the following section,
{\it the two conditions
(\ref{eq:condition1}) and (\ref{eq:condition2}),
and the positiveness of the structure function $\Phi(E,x)$}
suffice in order to
determine the energy spectrum of the quantum maximally
superintegrable systems.

There are only a few quantum 2-dimensional superintegrable systems
known. All the
examples studied in this paper have a classical counterpart.

\section{Examples of quantum superintegrable
\hfill\break two-dimensional systems}

In this section we shall apply the hypothesis of the previous
section in order to determine the energy spectrum of some quantum
superintegrable two-dimensional systems  using purely algebraic
methods.

\subsection{Harmonic oscillator in a flat space}

The  two-dimensional symmetric harmonic
oscillator in euclidean coordinates is described by the
hamiltonian:
\begin{equation}
H=\frac{1}{2}
\left( p_x^2 + p_y^2\right) +
 \frac{1}{2}\omega^2 \left( x^2 + y^2\right).
\label{eq:ho-hamiltonian}
\end{equation}

 The following Fradkin operators  \cite{Fradkin} can be defined:
\begin{equation}
B=S_{xx}-S_{yy} =
\left( p_x^2 + \omega^2 x^2\right) -
\left( p_y^2 + \omega^2 y^2\right),
\quad
S_{xy}= p_x p_y + \omega^2 xy.
\label{eq:Fradkin}
\end{equation}

The operator $B$ in (\ref{eq:Fradkin}) is the quantum mechanical analogue of
a third {\em constant of motion} in the sense of classical
hamiltonian mechanics \cite{Hiet1}, the
second one being the angular momentum operator:
\begin{equation}
L= xp_y- y p_x,
\label{eq:ang-mom}
\end{equation}
since:
\begin{equation}
\comm{H}{L}=\comm{H}{B}=0,
\label{eq:ho-comm-rel}
\end{equation}
while the operators $B,L$ do not commute, but they form a closed
algebra with the operator $S_{xy}$:
\begin{equation}
\comm{L}{B}=4 i S_{xy},
\quad
\comm{L}{S_{xy}}=-i B.
\label{eq:CommAlgebra}
\end{equation}

The above relations suggest the possibility of expressing the
two-di\-me\-n\-s\-io\-n\-al harmonic oscillator algebra by using the
deformed oscillator formulation:
\begin{equation}
\begin{array}{rl}
\displaystyle {\cal N}= \frac{L}{2}-u{\bf 1}, \\[0.2in]
\displaystyle {\cal A}^+ = \frac{B}{2}+ i S_{xy},  \\[0.2in]
\displaystyle {\cal A} = \frac{B}{2}- i S_{xy},
\end{array}
\label{eq:def-osc}
\end{equation}
where $u$ is a constant to be determined and
\begin{equation}
\begin{array}{rl}
\comm{\cal N}{\cal A^+}=& {\cal A^+}, \\
\comm{\cal N}{\cal A}=& -{\cal A}, \\
{\cal A}^+ {\cal A}  =&
H^2-\omega^2 (L-1)^2 \\
=&  H^2 -\omega^2 \left( 2 {\cal N} +2 u -1
\right)^2 \\
 = & \Phi ( H, {\cal N} ),
\end{array}
\label{eq:ho}
\end{equation}
where the function $\Phi (E,x)$ is given by:
\begin{equation}
\Phi(E,x)=E^2- \omega^2 \left( 2 x+2 u -1 \right)^2,
\label{eq:ho-str-function}
\end{equation}
and we can see that
$$
{\cal A} {\cal A}^+ = \Phi (H, {\cal N}+ {\bf 1} ).
$$

The existence of a finite dimensional representation of the
oscillator algebra is equivalent to the existence of a maximum
number $N+1$ which is a root of the structure function, with N being
the dimensionality of the algebra representation,  coinciding with
the dimensionality of the appropriate Fock space. This
restriction, combined with the annihilation of the structure function
for $x=0$,  is written as:
\begin{equation}
\begin{array}{c}
\Phi(E,0)=0, \\
\Phi(E,N+1)=0,\\
\Phi(E,x)>0\quad \mbox{for}\quad x=1,2,\ldots,N .
\end{array}
\label{eq:restrictions}
\end{equation}
Solving this system of two equations with two unknowns, $E$ and $u$, one
obtains the eigenvalues of the harmonic oscillator in a flat
space:
\begin{equation}
u= -\frac{N}{2}, \quad  E=E_N=\omega (N+1).
\label{eq:ho-energy}
\end{equation}
The angular momentum values allowed for each energy level can then
be determined by inserting the value of the constant $u$ just obtained
into the first of eq. (\ref{eq:def-osc}), the result being:
\begin{equation}
L=-N,-N+2,\ldots,N-2,N.
\label{eq:ho-ang-mom}
\end{equation}
The structure function of the deformed oscillator is calculated
to be:
$$
\Phi(E_N,x)=4 \omega^2 x ( N+1-x) .
$$
Clearly the above energy (\ref{eq:ho-energy})
and angular momentum spectra (\ref{eq:ho-ang-mom}) are
 the same as the ones
obtained by classical means, i.e. by solving the appropriate
Schr\"{o}dinger equation, or by using the SU(2) symmetry.
The existence of a finite dimensional algebra representation
should be attributed to the existence of stable periodical
trajectories in the corresponding classical case.
%
%

\subsection{Harmonic oscillator in a space with constant curvature}

Higgs \cite{Higgs} has studied the symmetries of a harmonic
oscillator in a non-flat space, a space with constant
curvature in particular. A typical example of such a space is the surface
of the sphere in a three dimensional space.

The curved space is geometrically described by the metric:
$$
ds^2 =
\frac{ dx^2 +dy^2 + \lambda (x dy -y dx)^2 }
     { \left( 1 + \lambda \left( x^2 +y^2\right) \right)^2},
$$
the flat space corresponding to $\lambda = 0$.
The harmonic oscillator in this space is defined in ref.
\cite{Higgs} by the Hamiltonian:
\begin{equation}
H=\frac{1}{2} \left(
\pi_x^2 + \pi_y^2 + \lambda L^2 \right) +
\frac{\omega^2}{2} \left( x^2 + y^2 \right),
\label{eq:Higgs-ho}
\end{equation}
where the angular momentum operator $L$ is given by   eq.
 (\ref{eq:ang-mom}) and
\begin{equation}
\begin{array}{c}
 \pi_x= p_x + \frac{\lambda}{2}
\left( x \left( x p_x + y p_y\right)+ \left( x p_x + y
p_y\right) x \right),\\
 \pi_y = p_y + \frac{\lambda}{2}
\left( y \left( x p_x + y p_y\right)+ \left( x p_x + y
p_y\right) y \right).
\end{array}
\label{eq:ext-momentum}
\end{equation}
By analogy to the harmonic oscillator in a flat space,
Higgs\cite{Higgs} has defined the Fradkin-like operators:
\begin{equation}
B=S_{xx}-S_{yy} =
\left( \pi_x^2 + \omega^2 x^2\right) -
\left( \pi_y^2 + \omega^2 y^2\right),
\label{eq:Higgs}
\end{equation}
\begin{equation}
S_{xy} =\frac{1}{2} \{ \pi_x, \pi_y \} + \omega^2 xy,
\end{equation}
which are symmetrized versions of eq. (\ref{eq:Fradkin}).

The commutators of $H$ with $L$ and $B$ are given in eq.
 (\ref{eq:ho-comm-rel}), while the operators
$B,L$ do not commute and their commutator is the same as in
equation (\ref{eq:CommAlgebra}), where the coordinates of the momentum $p$
should be replaced by the coordinates of the extended momentum
$\pi$, defined by eq. (\ref{eq:ext-momentum}) and already used in the
symmetrized Fradkin operators given above.

The operators $H,L,B,S_{xy}$ define again a closed non-linear algebra
as in the flat harmonic oscillator case.
Therefore we have another example of a superintegrable system in
a non-flat space.

In the present case we can also define the corresponding deformed oscillator
as in eq. (\ref{eq:def-osc}). The deformed algebra is then
completely defined by the structure function:
\begin{equation}
\begin{array}{rl}
\ &\Phi(E,x)= \  \\
 \ & E^2-
\left( \omega^2 + \frac{\lambda^2}{4} + \lambda E \right)
\left(2x+2u-1\right)^2 +
 \frac{\lambda^2}{4}\left( 2 x +2 u -1 \right)^4 .
\end{array}
\label{eq:c-ho-str-function}
\end{equation}

Following the same methodology as in the case of the harmonic
oscillator in a flat space,
 we can generate the corresponding Fock space for the
curved harmonic oscillator. By assuming then the existence of a
finite dimensional deformed algebra representation the
 restrictions corresponding to eq.
(\ref{eq:restrictions}) are valid. These equations determine
the energy eigenvalues:
\begin{equation}
E=E_N=\sqrt{ \omega^2 + \frac{\lambda^2}{4}} (N+1)
 + \frac{\lambda}{2}(N+1)^2,
\label{eq:curv-ho-energy}
\end{equation}
while the constant $u$ turns out to have the same values
as in eq. (\ref{eq:ho-energy}).
The angular momentum eigenvalues are again given by eq.
(\ref{eq:ho-ang-mom}).

Another interesting point arises from the comparison between the
structure function (\ref{eq:ho-str-function}) in a flat space
and eq. (\ref{eq:c-ho-str-function}) in a curved space: The geometry
of the space affects the algebra characterizing the harmonic oscillator.
For $\lambda= 0$ eq. (\ref{eq:c-ho-str-function}) reduces to eq.
(\ref{eq:ho-str-function}), as it should.

Finally, the structure function can be written as
$$
\Phi(E_N,x)=
4x(N+1-x)
\left(\lambda (N+1-x)+ \sqrt{\omega^2+\lambda^2/4}\right)
\left(\lambda x + \sqrt{\omega^2+\lambda^2/4}\right).
$$
The symmetries of the harmonic oscillator in a curved space have been
studied in ref. \cite{Z5} using the notion of the quadratic Racah algebras
QR(3).

%
%
\subsection{The Kepler problem in a curved space}

The case of the Kepler problem in a space with constant
curvature has been studied by Higgs \cite{Higgs}.
The hamiltonian is given by:
\begin{equation}
H=\frac{1}{2} \left(
\pi_x^2 + \pi_y^2 + \lambda L^2 \right) -
\frac{\mu}{r}, \quad
r=\sqrt{x^2+y^2},
\label{eq:Higgs-Kepler}
\end{equation}
where the angular momentum operator $L$ is given by   eq.
(\ref{eq:ang-mom}) and the $\pi_x,\pi_y$ are defined in eq.
(\ref{eq:ext-momentum}).

The Runge--Lenz vectors in the curved space can be defined by:
\begin{equation}
R_x=-\frac{1}{2}\acm{L}{\pi_y}+ \mu\frac{x}{r},
\quad
R_y=\frac{1}{2}\acm{L}{\pi_x}+ \mu\frac{y}{r}.
\label{eq:Runge-Lentz}
\end{equation}
This system is a quantum superintergrable system in a curved
space because:
$$
\comm{H}{L}=0, \quad \comm{H}{R_x}=0,
$$
and the operators $L$, $R_x$, $R_y$ form a closed algebra:
$$
\comm{L}{R_x}=iR_y, \quad \comm{L}{R_y}=-i R_x.
$$
Using the same hypothesis as previously we can define
the deformed oscillator algebra:
\begin{equation}
\begin{array}{rl}
{\cal N} =& L - u,  \\
{\cal A}^+ = & R_x +i R_y, \\
{\cal A} = & R_x - i R_y, \\
{\cal A}^+ {\cal A} = &
\mu^2 + 2 H \left (L - 1/2 \right)^2 -
           \lambda  \left (L - 1/2 \right)^2
           \left(  \left (L - 1/2 \right)^2 - 1/4 \right) \\
   =
\mu^2 +& 2 H \left ({\cal N}+u - 1/2 \right)^2 -
           \lambda  \left ({\cal N} +u - 1/2 \right)^2
           \left(  \left ({\cal N}+u - 1/2 \right)^2 - 1/4 \right) \\

   =& \Phi (H, {\cal N} ).
\end{array}
\label{eq:Kepler-algebra}
\end{equation}
The structure function in this case is defined by:
$$
\Phi(E,x)
=\mu^2 + 2 E \left (x+u - 1/2 \right)^2
$$
$$
     - \lambda  \left (x +u - 1/2 \right)^2
           \left(  \left (x+u - 1/2 \right)^2 - 1/4 \right).
$$
The solution of eqs (\ref{eq:restrictions}) is given by:
\begin{equation}
u=-\frac{N}{2}, \quad
E_N=-\frac{2 \mu^2}{(N+1)^2}+\lambda \frac{N(N+2)}{8}.
\label{eq:KeplerSpectrum}
\end{equation}
The permitted eigenvalues of the angular momentum operator $L$
are given by:
$$
L=-\frac{N}{2},-\frac{N}{2}+1,\ldots,\frac{N}{2}-1,\frac{N}{2}.
$$
This means that the symmetries of the Kepler problem are
compatible with the existence of angular momenta equal to 0,
1/2, 1, 3/2,\ldots. In  physical situations, however,
 only integer angular momenta appear, which
 means that $N=2n$. In this case the spectrum
given by eq.  (\ref{eq:KeplerSpectrum}) is the same to that
obtained by Higgs \cite{Higgs}. In the case of zero curvature, i.e.
$\lambda =0$, we obtain the usual Coulomb energy spectrum.

The structure function corresponding to the Kepler problem is given by:
$$
\Phi(E_N,x)= x (N+1-x)
\left(
 \frac{4 \mu^2}{(N+1)^2}+ \lambda \frac{(N+1-2x)^2}{4} \right).
$$
Zhedanov \cite{Z1} has proven that the nonlinear algebra of eq.
 (\ref{eq:Kepler-algebra}) can be approximated to second order by
the SU${}_q$(2) algebra \cite{Biedenharn,Macfarlane}.

The symmetries of the Kepler potential in a curved space have been
studied in ref. \cite{Z6} using the notion of the quadratic Racah algebras
QR(3).

\subsection{Fokas-Lagerstrom potential}

In classical mechanics the superintegrable system described by
the Hamiltonian:
\begin{equation}
H=\frac{1}{2}\left( p_x^2 + p_y^2 \right)
+\frac{x^2}{2}+\frac{y^2}{18}
\label{eq:FokasHamiltonian}
\end{equation}
has been studied by Fokas and Lagerstrom \cite{FL}. This
system has two additional classical invariants of motion,
\begin{equation}
J = p_x^2 + x^2, \quad \mbox{and} \quad
C=(x p_y - y p_x ) p_y^2 + \frac{y^3 p_x}{27}- \frac{xy^2 p_y}{3},
\label{eq:FokasClassIntegral}
\end{equation}
the second of which ($C$) is a cubic function of the coordinates.
The quantum version of the hamiltonian
(\ref{eq:FokasHamiltonian}) corresponds to a quantum
superintegrable system with two additional integrals:
$$J= p_x^2 + x^2,\quad
\mbox{and} \quad
B=\frac{1}{2}
\acm{x p_y - yp_x}{p_y^2} + \frac{y^3p_x}{27} -
\frac{\acm{x y^2}{p_y}}{6},
$$
where $\acm{\ }{\ }$ is the usual anticommutator. It is clear that
the quantum integral $B$ is the symmetrized version of the classical
integral $C$.
{}From the above definitions we can verify that:
$$ \cmm{H}{J}=0, \quad \cmm{H}{B}=0, $$
$$
\cmm{J}{B}=R, \quad \cmm{J}{R}=4 B,
$$
and
$$
\cmm{R}{B}=
8 J^3- 36J^2 H +48 J H^2 -16H^3 +\frac{56}{9} J - \frac{92}{9}H,
$$
$$
\begin{array}{rl}
R^2-4 B^2=& 4 J^4 -24 J^3 H +48 J^2 H^2 -32 J H^3 + \\
\ &+\frac{200}{9}J^2 -  \frac{616}{9}JH + 48 H^2 +\frac{20}{9}.
\end{array}
$$

{}From the above closed algebra we can define:
$$
{\cal N}= J/2-u, \quad
{\cal A}^+ = B +R/2, \quad
{\cal A} = B-R/2,
$$
where $u$ is a constant to be determined. These operators correspond
to a deformed oscillator algebra:
\begin{equation}
\begin{array}{rl}
\cmm{\cal N}{\cal A ^+} = & {\cal A}^+, \quad
\cmm{\cal N}{\cal A } = - {\cal A}, \\[0.2in]
{\cal A}^+ {\cal A} = & \displaystyle
\frac{1}{9} (J-1) (2H -J +1)(6H-3J+1)(6H-3J+5)\\[0.2in]
=&\displaystyle \frac{1}{9} (2 {\cal N}-1+2u) (2H -2u +1-2 {\cal
N}) \\[0.2in]
\ &  (6H-6u+1 -6{\cal N})(6H-6u+5-6{\cal N})\\[0.2in]
=&\Phi(H,{\cal N}), \\
{\cal A} {\cal A}^+ =& \Phi ( H, {\cal N}+1).
\end{array}
\end{equation}
The corresponding structure function is defined by:
$$
\begin{array}{rl}
\Phi (E,x) = & \frac{1}{9}(2 x-1+2u) (2 E -2u +1-2 x)\\
\ & (6 E -6u+1 -6x)(6 E -6u+5-6x).
\end{array}
$$

The existence of a finite representation of the algebra for each
energy eigenvalue implies that the structure function satisfies
eq.  (\ref{eq:restrictions}) and it is a positive function.
Therefore we can find the
possible energy eigenvalues having degeneracy equal to $N+1$:

Case \romannumeral 1 )
$$ u=1/2,\quad \mbox{and} \quad
E_N=N+1, $$
corresponding to the structure function:
$$
\Phi(E_N,x)=
16 x(N+1-x)\left( N+\frac{2}{3}-x\right)
\left( N + \frac{4}{3}-x \right).
$$

Case \romannumeral 2 )
$$ u=1/2,\quad \mbox{and} \quad
E_N=N+2/3 , $$
corresponding to the structure function:
$$
\Phi(E_N,x)=
16 x(N+1-x)\left( N+\frac{2}{3}-x\right)
\left( N + \frac{1}{3}-x \right).
$$

Case \romannumeral 3 )
$$ u=1/2,\quad \mbox{and} \quad
E_N=N+4/3 , $$
corresponding to the structure function:
$$
\Phi(E_N,x)=
16 x(N+1-x)\left( N+\frac{5}{3}-x\right)
\left( N + \frac{4}{3}-x \right).
$$

In all cases the degeneracy is determined by $J=2({\cal N} + u)$.
Since $\cal N$ obtains the $N+1$ values 0, 1, \dots, $N$, as a result
$J$ also obtains $N+1$ values.

The Hamiltonian (\ref{eq:FokasHamiltonian}) corresponds to the
linear combination of two harmonic oscillators which have the
above described energy spectrum.
 This case has a special significance, since it is an example of
a superintegrable potential which is not a separable one in two different
coordinate systems. The proposed method does not depend on the separability
of the variables in two systems.

\subsection{Smorodinsky--Winternitz potential}

The classical superintegrable Smorodinsky--Winternitz system
\hfill\break
\cite{SW1,SW2,SW3,Ev2} corresponds to the Hamiltonian:
\begin{equation}
H=\frac{1}{2}\left( p_x^2 + p_y^2 \right)
+k\left( x^2 + y^2\right) + \frac{c}{x^2}.
\label{eq:WinternitzHamiltonian}
\end{equation}
 This
system has two additional classical invariants of motion,
\begin{equation}
T= p_y^2 + 2 k y^2, \quad \mbox{and} \quad
C=x^2 p_y^2+ y^2 p_x^2 - 2 xy p_x p_y + 2 c \frac{y^2}{x^2},
\label{eq:WinternitzClassIntegral}
\end{equation}
the second of which ($C$) is a quartic function of the coordinates.
Evans \cite{Ev2} has proved that the
Winternitz--Smorodinsky potential in $N$ dimensions is an example of
a superintegrable system.
The quantum version of the hamiltonian
(\ref{eq:WinternitzHamiltonian}) corresponds to a quantum
superintegrable system with two additional integrals:
$$T = p_y^2 + 2ky^2,\quad
\mbox{and} \quad
B=x^2p_y^2+y^2p_x^2 -\acm{xy}{p_xp_y} + 2c\frac{y^2}{x^2}.
$$
It is clear that the quantum integral $B$ is the symmetrized version
of the classical integral $C$.
{}From the above definitions we can verify that:
$$ \cmm{H}{T}=0, \quad \cmm{H}{B}=0,  $$
and
$$
\cmm{T}{B}=R, \quad \cmm{T}{R}=32 k B + 8T^2-16 H T-16k,
$$
$$
\cmm{R}{B}=16 B T - 16 BH +32(c-1)T+8R+32H,
$$
$$
\begin{array}{rl}
R^2=&32kB^2+224kB+32(c-1)T^2+64HT+16RT-16RH-48H^2\\
\ &+16BT^2 -32BTH +192k(c-1) .
\end{array}
$$
{}From the above closed non-linear algebra we can define:
$$
\begin{array}{rl}
{\cal N}=& \frac{1}{\sqrt{32k}}T+u, \\
{\cal A}^+ =& 4 k B + \sqrt{\frac{k}{2}}R+ T^2 - 2HT-2k, \\
{\cal A} = &4 k B - \sqrt{\frac{k}{2}}R+ T^2 - 2HT-2k ,
\end{array}
$$
where $u$ is a constant to be determined. These operators correspond
to a deformed oscillator algebra:
\begin{equation}
\begin{array}{rl}
\cmm{\cal N}{\cal A ^+} = & {\cal A}^+, \quad
\cmm{\cal N}{\cal A } = - {\cal A}, \\
{\cal A}^+ {\cal A} = &
 24\,{H^2}\,k + 3\cdot {2^{{9\over 2}}}\,H\,{k^{{3\over 2}}} + 36\,{k^2} \\
\ &  - 96\,c\,{k^2} - {2^{{9\over 2}}}\,{H^2}\,{\sqrt{k}}\,T - 88\,H\,k\,T \\
\ &  - 3\cdot {2^{{9\over 2}}}\,{k^{{3\over 2}}}\,T +
  {2^{{{13}\over 2}}}\,c\,{k^{{3\over 2}}}\,T + 4\,{H^2}\,{T^2} \\
\ &+  3\cdot {2^{{7\over 2}}}\,H\,{\sqrt{k}}\,{T^2} + 44\,k\,{T^2} -
  16\,c\,k\,{T^2} \\
\ &  - 4\,H\,{T^3} - {2^{{7\over 2}}}\,{\sqrt{k}}\,{T^3} + {T^4}\\
=&\Phi(H,{\cal N}), \\
{\cal A} {\cal A}^+ =& \Phi ( H, {\cal N}+1).
\end{array}
\end{equation}
The corresponding structure function can be factorized as:
$$
\begin{array}{l}
\Phi (E,x) =
1024 k^2 \left(x-\left(u+\frac{3}{4}\right)\right)
\left(x-\left(u+\frac{1}{4}\right)\right) \\
 \left(\
x- \left( u+\frac{1}{2} + \frac{E}{\sqrt{8k}}
+\frac{\sqrt{1+8c}}{4} \right)
\right)  \left(\
x- \left( u+\frac{1}{2} + \frac{E}{\sqrt{8k}}
-\frac{\sqrt{1+8c}}{4} \right)
\right).
\end{array}
$$
The existence of a finite representation of the algebra for each
energy eigenvalue implies that the structure function satisfies
eq.  (\ref{eq:restrictions}). The positiveness of the
structure function for every $0<x\le N$ implies:
$$u=-\frac{3}{4},  $$
while  the energy eigenvalues are given by:
$$
E_N=\sqrt{8k}\left( N+ \frac{5}{4} + \frac{\sqrt{1+8c}}{4}
\right), \quad N=1,2,\ldots,
$$
with $-\frac{1}{8} \le c$.
The structure function is given by:
$$
\begin{array}{l}
\Phi(E_N,x)=\\
=1024 k^2 x \left( x+\frac{1}{2} \right)
\left( N+1 -x\right)
\left( N+1 + \frac{\sqrt{1+8c}}{2} -x \right).
\end{array}
$$

If the following restriction is valid:
\begin{equation}
 -\frac{1}{8}\le c \le \frac{3}{8},
\label{eq:Winter-restriction}
\end{equation}
the following energy eigenvalues are also permitted:
$$
E_N=\sqrt{8k}\left( N+ \frac{5}{4} - \frac{\sqrt{1+8c}}{4}
\right), \quad N=1,2,\ldots
$$
corresponding to the structure function:
$$
\Phi(E_N,x)=
1024 k^2 x \left( x+\frac{1}{2} \right)
\left( N+1 -x\right)
\left( N+1 - \frac{\sqrt{1+8c}}{2} -x \right) .
$$

In both cases the degeneracy of the levels is determined by
$T= \sqrt{32 k} ({\cal N} -u)$. Since $\cal N$ obtains the $N+1$
values 0, 1, \dots, $N$, as a result $T$ also obtains $N+1$ values.

It is worth noticing that in the case of solving the
problem using the Schr\"{o}\-dinger equation, the restriction
(\ref{eq:Winter-restriction}) is introduced by the assumption that the
eigenfunctions should be square integrable functions on the
plane $(x,y)$.
In the Schr\"{o}dinger equation solution the additional
restriction of finiteness of the potential energy restricts the choice
of $c$ to positive values only.

\subsection{The Holt potential}

The classical superintegrable Holt \cite{Holt} system
corresponds to the Hamiltonian:
\begin{equation}
H=\frac{1}{2}\left( p_x^2 + p_y^2 \right)
+\left( x^2 + 4y^2\right) + \frac{\delta}{x^2}.
\label{eq:HoltHamiltonian}
\end{equation}
This potential is a generalization of the harmonic oscillator potential
with a ratio of frequencies 2:1.
 This
system has two additional classical invariants of motion,
\begin{equation}
T= p_y^2 + 8 y^2, \quad \mbox{and} \quad
C=p_x^2 p_y + 8 xy p_x - 2 x^2 p_y + \frac{2 \delta}{x^2}p_y,
\label{eq:HoltClassIntegral}
\end{equation}
the second of them ($C$) being a cubic function of the momenta.
The quantum version of the hamiltonian
(\ref{eq:HoltHamiltonian}) corresponds to a quantum
superintegrable system with two additional integrals:
$$T = p_y^2 + 8y^2,\quad
\mbox{and} \quad
B=p_x^2 p_y + 4\acm{ xy}{ p_x} - 2 x^2 p_y + \frac{2 \delta}{x^2}p_y.
$$
It is clear that the quantum integral $B$ is the symmetrized version
of the classical integral $C$.
{}From the above definitions we can verify that:
$$ \cmm{H}{T}=0, \quad \cmm{H}{B}=0,  $$
and
$$
\cmm{T}{B}=R, \quad \cmm{T}{R}=32  B,
$$
$$
\cmm{R}{B} = -96 +256\delta -64 H^2
+128 HT -48 T^2,
$$
$$
R^2-32B^2=
1024H -704T +512\delta T -128T H^2 +128 T^2 H -32T^3.
$$
{}From the above closed non-linear algebra we can define:
$$
{\cal N}= \frac{T}{\sqrt{32}}-u, \quad
{\cal A}^+ = 8 B + \sqrt{2}R, \quad
{\cal A} = 8 B - \sqrt{2}R,
$$
where $u$ is a constant to be determined. These operators correspond
to a deformed oscillator algebra:
\begin{equation}
\begin{array}{rl}
\cmm{\cal N}{\cal A ^+} = & {\cal A}^+, \quad
\cmm{\cal N}{\cal A } = - {\cal A}, \\
{\cal A}^+ {\cal A} = &
2^6 \left(T-2\sqrt{2}\right)\\
\ &
\left(H-\frac{T}{2} +\sqrt{2} +\sqrt{\frac{1+8\delta}{2}}\right)
\left(H-\frac{T}{2} +\sqrt{2} -\sqrt{\frac{1+8\delta}{2}}\right)\\
=&\Phi(H,{\cal N}), \\
{\cal A} {\cal A}^+ =& \Phi ( H, {\cal N}+1).
\end{array}
\end{equation}
The corresponding structure function is defined by:
$$
\begin{array}{rl}
\Phi (E,x) =&
2^{\frac{23}{2}} \left((x+u)-\frac{1}{2}\right)\\
\ &
\left(\frac{E}{\sqrt{8}}-(x+u) +\frac{1}{2}
+\frac{\sqrt{1+8\delta}}{4}\right) \\
 & \left(\frac{E}{\sqrt{8}}-(x+u) +\frac{1}{2}
-\frac{\sqrt{1+8\delta}}{4}\right).
\end{array}
$$
The existence of a finite representation of the algebra for each
energy eigenvalue implies that the structure function satisfies
eq.  (\ref{eq:restrictions}). Therefore we can find the
possible energy eigenvalues having degeneracy equal to $N+1$:
$$ u=\frac{1}{2},\quad \mbox{and} \quad
E_N=\sqrt{8}\left(N+1 + \frac{\sqrt{1+8 \delta}}{4}\right),$$
where $(1+8\delta) \ge 0$.
The corresponding structure function is:
$$
\Phi(E_N,x)=
2^{\frac{23}{2}}
x(N+1-x)\left(N+1-x+ \frac{\sqrt{1+8 \delta}}{2}\right).
$$
In the special case where $-\frac{1}{8}\le \delta \le \frac{3}{8}$
there are energy eigenvalues given by:
$$ u=\frac{1}{2},\quad \mbox{and} \quad
E_N=\sqrt{8}\left(N+1 - \frac{\sqrt{1+8 \delta}}{4}\right),$$
and the structure function is:
$$
\Phi(E_N,x)=
2^{\frac{23}{2}}
x(N+1-x)\left(N+1-x - \frac{\sqrt{1+8 \delta}}{2}\right),
$$
which is positive for $0<x\le N$ if $-\frac{1}{8}\le\delta\le \frac{3}{8}$.

In both cases the degeneracy of the levels is determined by
$T= \sqrt{32} ({\cal N} +u )$. Since $\cal N$ is obtaining the $N+1$ values
0, 1, \dots, $N$, as a result $T$ also obtains $N+1$ values.
The quantum Holt potential has also been studied recently by using quadratic
algebras by L\'{e}tourneau and Vinet \cite{LV}.

\section{Discussion}

In this paper, starting from  classical superintegrable
systems, we have shown
that the corresponding quantum systems are superintegrable ones,
the quantum integrals (quantum constants of motion)
being obtained from the classical ones using a  symmetrization procedure.
Furthermore, the quantum superintegrable
systems can be  described in terms of a  deformed oscillator algebra.
The operators of the deformed oscillator algebra are constructed from
the quantum integrals. The deformed oscillator algebra is characterized
by a structure function $\Phi(E,N)$, which takes a specific form for
each superintegrable system. The eigenvalues of the energy
and their degeneracies  are determined
in an economical way
directly from equations satisfied by the structure function, the results
being in agreement with these coming from the independent solution of the
relevant Schr\"odinger equation.

A few comments and some open problems are now in place:

i) In all of the examples considered in this paper, quantum
superintegrability is induced by classical superintegrability, the
quantum integrals of motion being symmetrized versions of the corresponding
classical integrals. The extend to which classical superintegrability
implies in general quantum superintegrability as well, has to be tested.

ii) In all of the examples considered in this paper, quantum
superintegrability manifests itself in the degeneracy of the energy
levels, a fact already noticed \cite{KW1}.

iii) The hypothesis that to each superintegrable system corresponds a
deformed oscillator algebra, i.e.
\begin{center}
superintegrability $\rightarrow$ deformed oscillator algebra
\end{center}
is an exact proposition in the classical case, as shown in section 2.
In the quantum case, however, a general formal proof is still lacking. In
 section 3 a working hypothesis was made, which was proved successful in the
examples considered in section 4.

iv) The list of two-dimensional quantum superintegrable systems given in
this paper is not exhaustive. There are classical superintegrable systems
for which the quantum superintegrability has to be proven, as, for
example, the Calogero system (\cite{Hiet1}, eq. (3.5.9)), which possesses
a sixth order invariant. Furthermore, there are two-dimensional
systems for which the quantum superintegrability has been shown, but
the determination of the corresponding deformed oscillator algebra
requires heavy computation, as, for example, the Winternitz--Smorodinsky
potential of ref. \cite{SW2}, given also in \cite{Hiet1}, eq. (3.2.36).

v) The extension of the present method to three-dimensional quantum
superintegrable systems is under investigation. It should be mentioned that
classical superintegrable systems in 3 dimensions having invariants
which are quadratic polynomials in the canonical momenta have been
recently studied in \cite{Ev2}.

vi) Another interesting point is the semiclassical study of
two-di\-me\-n\-s\-i\-o\-nal
superintegrable systems. The algebra characterizing the
two-di\-me\-n\-s\-i\-o\-nal
classical superintegrable systems, studied in section 2, can be
quantized by using the correspondence \{ Poisson bracket $\to$
commutator \}. Through this procedure, from the classical algebra
a quantum deformed algebra is obtained, which is the semiclassical
counterpart of the exact quantum oscillator algebra considered in this
paper, these two algebras being slightly different.

vii) The example of the Fokas Lagerstrom potential, which is the oscillator
with ratio of frequencies 1:3, shows that quantum superintegrability
implies a dynamical symmetry. This example was examined
using algebraic methods for the first time. All the other examples have been
already studied by other authors, as it has been indicated in the text.
The difference of the proposed treatment is that we do not use the
separability in two coordinate systems in order to calculate the dynamical
symmetries. The set of  two-di\-me\-n\-sio\-nal systems separable in
two different coordinate systems is a subset of the class of
 the superintegrable systems in two dimensions.
The study of other superintegrable systems non-separable in more than one
coordinate system seems to be very interesting. A class of such
systems already known consists of the oscillators with rational
ratio of frequencies.

viii) Many of these examples
\cite{LV,Z5,Z6} have been studied by using cases of
quadratic Askey--Wilson algebras QAW(3) \cite{Z4}. An open problem is if
the general quadratic Askey--Wilson algebra can be expressed by
a deformed oscillator. One can also notice that the algebra of the
Fokas--Lagerstrom problem is a cubic algebra, while all the other
examples in this paper correspond to quadratic algebras.

\bigskip
{\bf Acknowledgment}

One of the authors (DB) is grateful to the Greek Ministry of Research
and Technology for support.

\vfill\eject


\begin{thebibliography}{99}

\bibitem{MQL}
M.~Moshinsky, C.~Quesne and G.~Loyola, Ann. Phys. (NY) {\bf 198}, 103 (1990).

\bibitem{Hiet1}
J. Hietarinta, Phys. Rep. {\bf 147}, 87 (1987).

\bibitem{Ev}
N. W. Evans, Phys. Rev. A {\bf 41}, 5666 (1990).

\bibitem{KW1}
M. Kibler and P. Winternitz, Phys. Lett. A {\bf 147}, 338 (1990).

\bibitem{Higgs}
P. W. Higgs,  J. Phys. A {\bf 12}, 309 (1979).

\bibitem{Leemon}
H.~I.~Leemon, J. Phys. A {\bf 12}, 489 (1979).

\bibitem{FL}
A. S. Fokas and P. A. Lagerstrom, J. Math. Anal. Appl. {\bf 74}, 325 (1980).

\bibitem{SW1}
P. Winternitz, Ya. A. Smorodinsky, M. Uhlir and I. Fris,
Yad. Fiz. {\bf 4}, 625 (1966) [Sov. J. Nucl. Phys.
{\bf 4}, 444 (1966)].

\bibitem{SW2}
I. Fris, V. Mandrosov, Ya. A. Smorodinsky, M. Uhlir and P. Winternitz,
Phys. Lett. {\bf 16}, 354 (1965).

\bibitem{SW3}
A. A. Makarov, Ya. A. Smorodinsky, K. Valiev and P. Winternitz, Nuovo
Cimento {\bf 52}, 1061 (1967).

\bibitem{Ev2}
N.~W.~Evans, Phys. Lett. A {\bf 147}, 483 (1990).
(1990) 483

\bibitem{Holt}
C. R. Holt, J. Math. Phys. {\bf 23}, 1037 (1982).

\bibitem{H1}
H. Hartmann, Theor. Chim. Acta {\bf 24}, 201 (1972).

\bibitem{H2}
H. Hartmann, R. Schuck and J. Radtke, Theor. Chim. Acta {\bf 42}, 1 (1976).

\bibitem{H3}
H. Hartmann and D. Schuch, Int. J. Quant. Chem. {\bf 18}, 125 (1980).

\bibitem{KW2}
M. Kibler and P. Winternitz, J. Phys. A {\bf 20}, 4097 (1987).

\bibitem{Weigert}
S. Weigert, Physica D {\bf 56}, 107 (1992).

\bibitem{Robnik}
M. Robnik, J. Phys. A {\bf 19}, L841 (1986).

\bibitem{Korsch}
H. J. Korsch, Phys. Lett. A {\bf 90}, 113 (1982).

\bibitem{Hiet2}
J. Hietarinta, J. Math. Phys. {\bf 25}, 1833 (1984).

\bibitem{KR}
P. P. Kulish and N. Yu. Reshetikhin, Zapiski Semenarov LOMI {\bf 101}, 101
(1981).

\bibitem{Sklyanin}
E. K. Sklyanin, Funct. Anal. Appl. {\bf 16}, 262 (1982).

\bibitem{Drinfeld}
V. G. Drinfeld, in {\it Proceedings of the International Congress
 of Mathematicians},
ed. A. M. Gleason (American Mathematical Society, Providence, RI, 1986) p. 798.

\bibitem{Jimbo}
M. Jimbo, Lett. Math. Phys. {\bf 11}, 247 (1986).

\bibitem{Jimbo2}
M. Jimbo, {\it Yang Baxter Equation in Integrable Systems}, Adv. Series
in Math. Phys. {\bf 10} (World Scientific, Singapore, 1990).

\bibitem{Z1}
 A.~S.~Zhedanov, Mod. Phys. Lett. A {\bf 7}, 507 (1992).

\bibitem{Z2}
Ya. I. Granovskii, A. S. Zhedanov and I. M. Lutzenko, J. Phys. A {\bf 24},
3887 (1991).

\bibitem{Z3}
O.~F.~Gal'bert, Ya.~I.~Granovskii and A.~S.~Zhedanov,
Phys. Lett. A {\bf 153}, 177 (1991).

\bibitem{Z4}
Ya. I. Granovskii, I. M. Lutzenko and A. S. Zhedanov,
Ann. Phys. (NY) {\bf 217}, 1 (1992).

\bibitem{LV}
P. L\'{e}tourneau and L. Vinet,
{\it Quadratic algebras in Quantum Mechanics},
Univ. of Montr\'{e}al preprint UdeM-LPN-TH-93-13.

\bibitem{KC}
M.~Kibler and C.~Campigotto, IPN de Lyon preprint LYCEN 9204 (March 1992).

\bibitem{PRAnote}
D. Bonatsos, C. Daskaloyannis and K. Kokkotas,
Phys. Rev. A, in press.


\bibitem{Biedenharn}
 L.~C.~Biedenharn, J. Phys. A {\bf 22}, L873 (1989).

\bibitem{Macfarlane}
A.~J.~Macfarlane, J. Phys. A {\bf 22}, 4581 (1989).

\bibitem{AC}
M.~Arik and D.~D.~Coon, J. Math. Phys. {\bf 17}, 524 (1976).

\bibitem{Ku}
M. V. Kuryshkin,  Ann. Fond. Louis de Broglie {\bf 5}, 111 (1980).

\bibitem{JBM}
A. Jannussis, G. Brodimas and R. Mignani, J. Phys. A {\bf 24}, L775 (1991).

\bibitem{CJ}
R. Chakrabarti and R. Jagannathan, J. Phys. A {\bf 24}, L711 (1991).

\bibitem{OK}
 Y.~Ohnuki and S.~Kamefuchi, {\it Quantum Field Theory
and Parastatistics} (Springer Verlag, Berlin, 1982).

\bibitem{FV}
 R.~Floreanini and L.~Vinet,  J.Phys. A {\bf 23}, L1019 (1990).

\bibitem{OKK}
K. Odaka, T. Kishi and S. Kamefuchi, J. Phys. A {\bf 24}, L591 (1991).

\bibitem{VPJ}
 K.~S.~Viswanathan, R.~Parthasarathy and
R.~Jagannathan, J. Phys. A {\bf 25}, L335 (1992).

\bibitem{Hayashi}
T, Hayashi, Commun. Math. Phys. {\bf 127}, 129 (1990).

\bibitem{BD1}
D. Bonatsos and C. Daskaloyannis, J. Phys. A {\bf 26}, in press (1993).

\bibitem{D1}
 C.~Daskaloyannis, J. Phys. A {\bf 24}, L789 (1991).

\bibitem{JBSZ}
 A.~Jannussis, G.~Brodimas, D.~Sourlas and V.~Zisis,
Lett. Nuovo Cimento {\bf 30}, 123 (1981) 123

\bibitem{BJ}
G. Brodimas, A. Jannussis, D. Sourlas, V. Zisis and P. Poulopoulos,
Lett. Nuovo Cimento {\bf 31}, 177 (1981).

\bibitem{BD}
 J.~Beckers and N.~Debergh, J. Phys. A {\bf 24}, L1277 (1991).

\bibitem{AD}
M.~Arik, E.~Demircan, T.~Turgut, L.~Ekinci and M.~Mungan,
Z. Phys. C {\bf 55}, 89 (1992).

\bibitem{FN}
D.~Fairlie and J.~Nuyts, Z. Phys. C {\bf 56}, 237 (1992).

\bibitem{D2}
C.~Daskaloyannis, {\it J. Phys. A} {\bf 25}, 2261 (1992).

\bibitem{D3}
 C.~Daskaloyannis and K.~Ypsilantis, J. Phys. A {\bf 25}, 4157 (1992).

\bibitem{BD2}
D. Bonatsos and C. Daskaloyannis, Chem. Phys. Lett. {\bf 203}, 150 (1993).

\bibitem{Fradkin}
 D.~M.~Fradkin, Am. J. Phys. {\bf 33}, 207 (1965).

\bibitem{Z5}
Ya. I. Granovskii, A. S. Zhedanov and I. M. Lutzenko,
Teor. Mat. Fiz. {\bf 91}, 207 (1992) (in Russian).

\bibitem{Z6}
Ya. I. Granovskii, A. S. Zhedanov and I. M. Lutzenko,
Teor. Mat. Fiz. {\bf 91}, 396 (1992) (in Russian).

\end{thebibliography}
\end{document}